# Impact of the activation rate of the hyperpolarization-activated current I$_h$ on the neuronal membrane time constant and synaptic potential duration

Cesar C. Ceballos[1,2], Rodrigo F.O. Pena[1,3], and Antonio C. Roque[1*]

[1]Department of Physics, School of Philosophy, Sciences and Letters of Ribeirão Preto, University of São Paulo, Ribeirão Preto, SP, Brazil. [2]Current address: Vollum Institute, Oregon Health & Science University, Portland, OR, USA. [3]Current address: Federated Department of Biological Sciences, New Jersey Institute of Technology and Rutgers University, Newark, New Jersey, NJ, USA.

**Abstract**
The temporal dynamics of membrane voltage changes in neurons is controlled by ionic currents. These currents are characterized by two main properties: conductance and kinetics. The hyperpolarization-activated current (I$_h$) strongly modulates subthreshold potential changes by shortening the excitatory postsynaptic potentials and decreasing their temporal summation. Whereas the shortening of the synaptic potentials caused by the I$_h$ conductance is well understood, the role of the I$_h$ kinetics remains unclear. Here, we use a model of the I$_h$ current model with either fast or slow kinetics to determine its influence on the membrane time constant ($\tau_m$) of a CA1 pyramidal cell model. Our simulation results show that the I$_h$ with fast kinetics decreases $\tau_m$ and attenuates and shortens the excitatory postsynaptic potentials more than the slow I$_h$. We conclude that the I$_h$ activation kinetics is able to modulate $\tau_m$ and the temporal properties of excitatory postsynaptic potentials (EPSPs) in CA1 pyramidal cells. In order to elucidate the mechanisms by which I$_h$ kinetics controls $\tau_m$, we propose a new concept called "time scaling factor". Our main finding is that the I$_h$ kinetics influences $\tau_m$ by modulating the contribution of the I$_h$ derivative conductance to $\tau_m$.

## 1. INTRODUCTION

Hyperpolarization-activated cyclic nucleotide–gated (HCN) channels are non-selective cationic channels that underlie the hyperpolarization-activated cation current (I$_h$). These channels are the main controllers of synaptic integration at the distal dendrites in cortical and hippocampal neurons, where they are mostly expressed [1-3]. Similarly to K$^+$ channels, the HCN channel is a tetramer, made of the different combinations of subunits 1-4 [4,5], which leads to a diverse set of kinetics [4,6,7]. The subunit HCN1 presents the fastest kinetics of all four [4,6,7,8], and down regulation of HCN1 channels results in an increase in postsynaptic excitability [5,9,10,11,12,13,14,15].

The HCN1 and HCN2 isoforms are predominantly expressed in the dendrites of cortical and hippocampal pyramidal neurons. They are particularly suited to attenuate and shorten the excitatory postsynaptic potentials (EPSPs) of those cells [2]. Accordingly, pharmacological inhibition of I$_h$ was found to increase summation of EPSPs [9,16,17] and genetic HCN1 deletion has been shown to be associated to lengthening of EPSPs in cortical and hippocampal pyramidal neurons [18,19]. Specific loss of the fast HCN1 subunit reduces I$_h$ and slows down the activation/deactivation kinetics of I$_h$, prolonging the EPSP decay and increasing their temporal summation window [2,14,15,20,21].

The membrane time constant ($\tau_m$) reflects the charging of the membrane capacitor and is traditionally thought to be determined only by the conductance of the subthreshold currents (linear and voltage dependent). In a purely passive membrane, the leak conductance determines $\tau_m$ in a manner such as the bigger the leak conductance the smaller $\tau_m$. However, neurons also express voltage dependent currents with specific activation/deactivation kinetics providing them with a variety of dynamical temporal adjustments [4,22,23]. Surprisingly, the effect of the current kinetics has been long neglected and thought not connected to $\tau_m$. Although Koch et al. [23] proposed that voltage-dependent currents would be able to modulate $\tau_m$ in a more complex manner than classical passive currents, there were no recent significant advances in the understanding of how I$_h$ kinetics affects $\tau_m$.

The activation time constant of I$_h$ ($\tau_h$) lies in a range from tens of milliseconds to several seconds and is well fitted by the sum of two exponentials with fast and slow time constants [4]. This might be related to the expression of at least two different subunits of HCN channels such as HCN1 and HCN2 in cortical neurons [4,24,25,26]. Here we ask whether and how the fast and slow kinetics of I$_h$ contributes to $\tau_m$ as well as how these contributions would differ. In the positive case, we hypothesize that the faster activation would decrease $\tau_m$ more strongly than the slow activation. For this, we use an $I_h$ current model with either fast or slow $\tau_h$ embedded in a single compartment CA1 pyramidal neuron model. The model is composed of a leak current and the $I_h$ current. Our results show that $I_h$ with fast $\tau_h$ decreases $\tau_m$ and shortens artificial EPSPs in a stronger manner than $I_h$ with slow $\tau_h$. Furthermore, we also elucidate the mechanisms by which $I_h$ with fast kinetics decreases more $\tau_m$ than the slow $I_h$. In order to elucidate the biophysical mechanisms underlying the influence of $\tau_h$ on $\tau_m$, we propose a new analytical solution of the $\tau_m$ as an extension of the definition of $\tau_m$ from the classical passive cable theory.

---

[*] antonior@usp.br

2. **METHODS**

**2.1. Neuron model**

In our model, we consider a single compartment neuron model where the membrane voltage is described by the equation

$$C \frac{dV}{dt} = -I_h - I_L + I(t), \tag{1}$$

where $C$ is the membrane capacitance, $I_h$ is the hyperpolarization-activated current, $I_L$ is the leak current, and $I(t)$ is an external current. The $I_h$ current is modeled using the Hodgkin-Huxley formalism as:

$$I_h = \overline{g_h} A_h(V,t)(V - E_h), \tag{2}$$

where the maximum conductance $\overline{g_h}$ is in nS and the reversal potential $E_h$ is in mV.

The activation variable $A_h$ is represented by the Boltzmann term [27]:

$$\frac{dA_h(V,t)}{dt} = \frac{A_h^\infty - A_h(V,t)}{\tau_h}, \tag{3}$$

where $\tau_h$ is the activation time constant in ms and $A_h^\infty$ is the steady state activation variable described by

$$A_h^\infty = \frac{1}{1 + e^{(V - V_{1/2})/k}}. \tag{4}$$

In Eq. 4 we used $V_{1/2} = -82$ mV and $k = 9$ mV [2]. The leak conductance is modeled by the equation $I_L = \bar{g}_L(V - E_L)$ where $\bar{g}_L$ is the maximum conductance in nS and $E_L$ is the reversal potential in mV.

The input conductance $G_{in}$, or *slope* conductance (since it corresponds to the slope of the steady-state *I-V* relation) of the model is obtained by differentiating the total current in Eq. 1 with respect to $V$ in the steady state, i.e., when $dV/dt = 0$. Since the total current in this case is $I = I_h + I_L$, we have

$$G_{in} = \frac{dI}{dV} = G_h + g_L, \tag{5}$$

where $G_h$ is the slope conductance of the $I_h$ current and $g_L$ is the leak conductance.

The $I_h$ slope conductance is obtained differentiating the current in Eq. 2

$$G_h = \frac{dI_h}{dV_{ss}} = \overline{g_h} A_h^\infty + \overline{g_h}(V - E_h)\frac{d(A_h^\infty)}{dV_{ss}}, \tag{6}$$

where the first term is the chord conductance ($g_h$) and the second term is the derivative conductance ($G_h^{Der}$) [28,29,30]. The derivative of $A_h^\infty(V)$ in Eq. 6 can be obtained differentiating Eq. 4: $\frac{dA^\infty}{dV} = \frac{(A^\infty - 1)A^\infty}{k}$, which implies that the derivative conductance is:

$$G_h^{Der} = \overline{g_h} A_h^\infty \frac{(A_h^\infty - 1)}{k}(V - E_h). \tag{7}$$

Notice that the chord conductance ($g_h = I_h/(V - E_h)$) is always positive, whereas $G_h^{Der}$ can be positive or negative.

The single compartment is a cylinder with length and basis diameter 70 µm. The maximum conductance and reversal potential of the $I_h$ and $I_L$ are, respectively $\overline{g_h}$ = 10 nS and $E_h$ = −30 mV; and $\overline{g_L}$ = 10 nS and $E_L$ = −90 mV. In our simulations of the model, the initial membrane potential was −90 mV and the time step was 0.1 ms. Simulations were run in the NEURON simulator [31] and the results were analyzed in MATLAB (THE MATHWORKS, Natick, MA). Table 1 summarizes the nomenclature used to represent the conductances and Fig 1A shows their numerical properties. Also Fig 1B and Fig 1C shows the numerical properties of the steady state activation variable $A_h^\infty$.

| Symbol | Definition |
|---|---|
| $\overline{g_L}$ | Leak maximum conductance |
| $\overline{g_h}$ | $I_h$ maximum conductance |
| $g_h$ | $I_h$ chord conductance |
| $G_h^{Der}$ | $I_h$ derivative conductance |
| $G_h$ | $I_h$ slope conductance |

**Table 1. Conductances appearing in the model.**

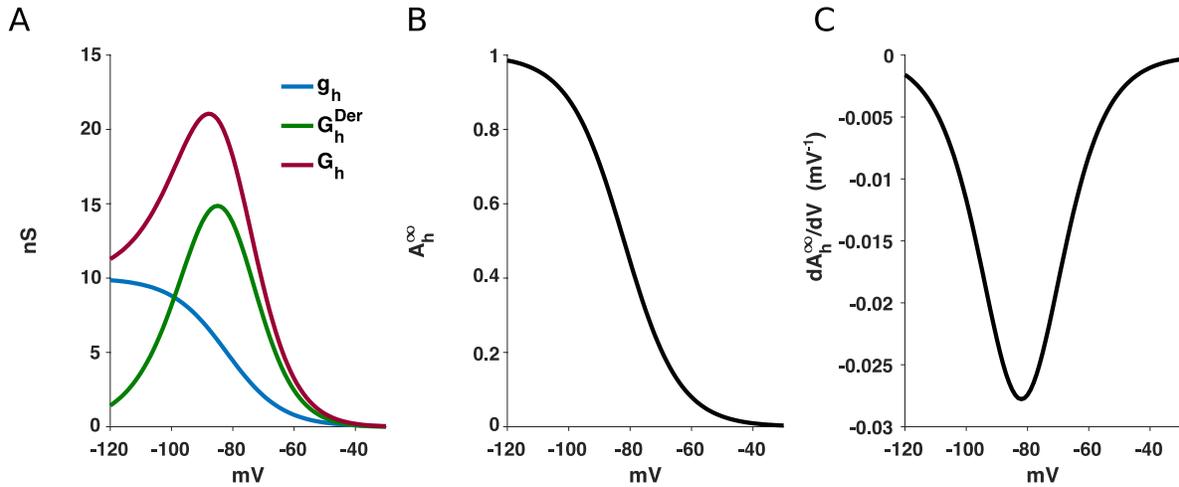

**Fig 1.** Voltage-dependent properties of the $I_h$. A. Slope conductance ($G_h$), chord conductance ($g_h$), and derivative conductance ($G_h^{Der}$). B. Steady state activation variable ($A_h^\infty$). C. Voltage derivative of the steady state activation variable ($\frac{dA_h^\infty}{dV}$).

### 2.2. Simulation data analysis

The membrane time constant ($\tau_m$) was determined from computational simulations analyzing voltage-current (V-I) relationships. V-I curves in current-clamp were obtained by injecting 4 s current pulses with +10 pA steps from −100 pA to 300 pA. Membrane time constant was measured by first applying 4 s long current to achieve the desired test potential and then injecting an additional small current step (+20 pA, 4 s) and fitting a single exponential to the initial rise of the voltage. Since simulations with $I_h$ include sag, we fitted only from the initial rise voltage until the moment the maximum value was reached. Moreover, the effect of $I_h$ on the amplitude and duration of EPSPs was determined by injecting a depolarizing EPSC at the end of the trace.

We also inject sinusoidal currents with linearly increasing frequencies (ZAP protocol) as in [29]. For that, the current $I = A\,sin[\pi(f(t) - F_{start})(t - t_{start})]$ has its frequency described by $f(t) = F_{start} + \frac{(F_{stop} - F_{start})(t - t_{start})}{t_{stop} - t_{start}}$ where $F_{start}$ ($F_{stop}$) is the initial (final) input frequency and $t_{start}$ ($t_{stop}$) marks the initial (final) moment of current injection. Below we convert frequencies to $\omega$, where $\omega = 2\pi f$. Standard values are A =10 pA, $F_{start}$ = 0.001 Hz and $F_{stop}$ = 40 Hz, $t_{start}$ = 1 s and $t_{stop}$ = 10 s.

Data were analyzed using routines written in Matlab (THE MATHWORKS, Natick, MA) and figures were generated in GraphPad Prism (GRAPHPAD SOFTWARE, La Jolla, CA).

## 3. RESULTS

### 3.1. $I_h$ kinetics affects EPSP attenuation

The main goal of this work was to determine the effect of the activation kinetics on the EPSP shape. It is well known that $I_h$ attenuates and shortens EPSPs [2], however it still remains unknown how the $I_h$ kinetics influences this attenuation. In order to test the effect of the $I_h$ activation kinetics on the EPSP shape, we ran simulations where a fixed EPSC was applied in different membrane potentials. We ran simulations in a neuron with a single compartment with a leak current and $I_h$ using two $\tau_h$ values: a fast one of 10 ms and a slow one of 500 ms. As expected, simulated EPSPs amplitude and area are decreased with hyperpolarized membrane potentials. This is caused by the increased $g_h$ and $G_h^{Der}$ due to hyperpolarization as it has been discussed elsewhere [29]. Interestingly, all the EPSP parameters were smaller when $\tau_h$ was faster (Fig 2B, 2C and 2D). These results suggest that the kinectis of $I_h$ has a strong impact on the EPSP shape, with the $I_h$ with fast activation kinetics being more efficient in attenuating the EPSPs than the $I_h$ with slower kinectics.

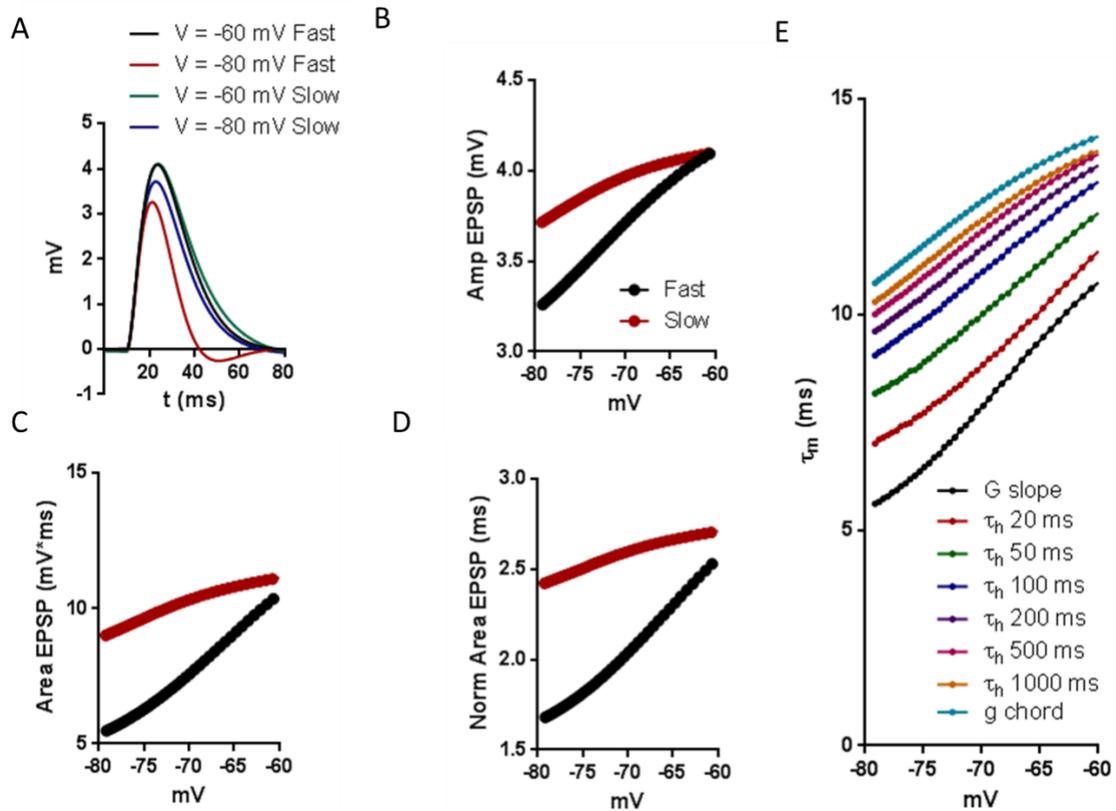

**Fig 2.** A. simulated EPSP representatives for membrane potentials -60 and -80 mV and for fast and slow $\tau_h$. B. simulated EPSP amplitude obtained with $I_h$ with fast activation ($\tau_h$ = 10 ms) and slow activation ($\tau_h$ = 500 ms) for different membrane potentials. C. EPSP area D. EPSP area normalized by the amplitude. E. Computational simulations for a single compartment with a leak current and $I_h$. Voltage dependent effect on

$\tau_m$ of $\tau_h$ with activation kinetics of 20, 50, 100, 200, 500 and 1000 ms. We also calculated $\tau_m = CR_{in}$ (G slope) and $\tau_m = C/(\bar{g}_L + g_h)$ (g chord).

The EPSP decay is mainly determined by $\tau_m$ [28]. In Fig 2A-D, we observe that $I_h$ with fast kinetics shortens the EPSP decay more than $I_h$ with slower kinetics. Based on these results, we hypothesize that $I_h$ activation kinetics ($\tau_h$) might influence the EPSP decay by influencing $\tau_m$. In order to study the effect of $\tau_h$ on $\tau_m$, we ran simulations of a single compartment neuron with a leak current and $I_h$, and then varied $\tau_h$ from 20 ms to 1000 ms. Fig 2E shows the voltage dependence of $\tau_m$ for different values of $\tau_h$. As expected, $\tau_m$ decreases with hyperpolarization, i.e., $\tau_m$ decreases in the same direction of $I_h$ activation. Furthermore, for a given membrane potential, $\tau_m$ decreases when $\tau_h$ decreases, suggesting that increasing the activation rate of $I_h$ will consequently increase the rate of change of membrane potential. We can then conclude that $I_h$ kinetics shortens the EPSPs by decreasing $\tau_m$. In addition, notice that for any value of $\tau_h$, $\tau_m$ is restricted to some limits, i.e. for $\tau_h \to 0$ we have $\tau_m = CR_{in}$ and for $\tau_h \to \infty$ we have $\tau_m = C/(g_L + g_h)$, where $R_{in}$ is the steady state input resistance. Thus, we conclude that maximum EPSP attenuation is reached when $\tau_h = 0$ and the minimum when $\tau_h \to \infty$.

$I_h$ modulation of the neuron temporal dynamics strongly depends on the voltage change speed [29]. As an example, note that EPSP amplitude and area curves (Figs. 2B-D) get closer for depolarized voltages than $\tau_m$ curves (Fig 2E). This is owing to the fact that $I_h$ behaves as a high-pass filter causing a much stronger effect in voltage changes that evolve slowly (as the voltage changes used to measure $\tau_m$) compared to faster events as EPSPs. Also, this effect is stronger when $I_h$ is more activated (see Fig 1B).

Moreover, similar normalized areas of EPSPs were obtained under similar $\tau_m$ values (e.g. ~2.4 for normalized area of EPSPs at -80 mV for slow and at -63 mV for fast $\tau_h$ (Fig 2D) and their corresponding $\tau_h$ values ($\tau_m \approx$ 10 ms, see Fig 1E)). This suggests that the observed changes in EPSPs temporal properties are likely caused by changes in $\tau_m$. To confirm this, we ran simulations by changing $g_L$, $g_h$ and the capacitance while computing $\tau_m$ and the normalized EPSP area.. Fig 3 shows good linear fits between $\tau_m$ and the normalized EPSP area, suggesting that changes in $\tau_m$ will lead to changes in EPSP temporal properties, regardless of the source of this variation (i.e., changes in conductance or capacitance).

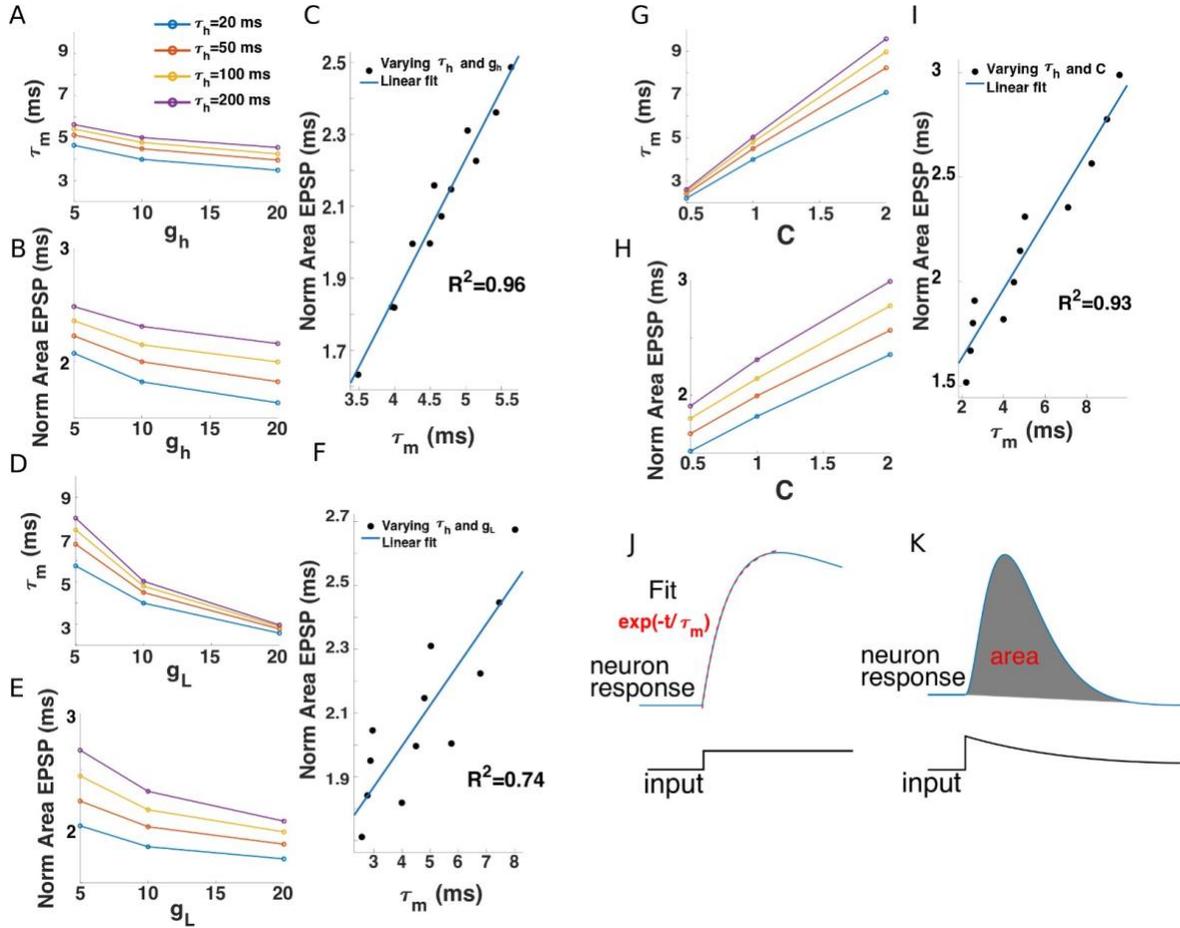

**Fig 3.** EPSP shortening is mediated by $\tau_m$. (A,D,G) $\tau_m$ as function of $g_h$ (in A), $g_L$ (in E), C (in G), and $\tau_h$ (in all). (B,E,H) EPSP area normalized by the amplitude as function of $g_h$ (in B), $g_L$ (in E), C (in H), and $\tau_h$ (in all). (C,F,I) Linear fit of normalized area plotted against $\tau_m$ for each of the cases where $g_h$, $g_L$, and C are respectively varied with $\tau_h$. Values of $R^2$ can be found in the plots. (J) $\tau_m$ was obtained by fitting the voltage change driven by square input current as a single exponential function. (K) EPSP area was calculated as the area under the trace.

### 3.2. Time scaling factor mediates modulation of the membrane time constant by rate of activation of $I_h$

In the previous section, by using computer simulations we established that $I_h$ with fast kinetics shortens the EPSP decay more efficiently than $I_h$ with slower kinetics, and that this is caused by an effect on $\tau_m$. However, the biophysical mechanism underlying this influence of $\tau_h$ on $\tau_m$ remains unclear. To shed some light on this mechanism, we introduce a new concept, which we will call "time scaling factor", or simply α factor.

The concepts of chord conductance and slope conductance are fundamental to explain the effect of $I_h$ with infinitely slow or instantaneous activation, i.e., when $\tau_h \to \infty$ or $\tau_h \to 0$, on $\tau_m$. For instance, in a single compartment neuron with a leak current ($I_L$) and $I_h$ current, when $\tau_m(\tau_h \to \infty) = \frac{C}{\bar{g}_L + g_h}$, on the other hand, when $\tau_h = 0$, $\tau_m = \frac{C}{\bar{g}_L + G_h}$, where $g_h$ is the $I_h$ chord conductance and $G_h$ is the $I_h$ slope conductance (see Appendix A, Eq. A9 and A10, also see [22]). In summary, $\tau_m$ is determined by the steady state slope conductance of the instantaneous current and chord conductance of the infinitely slow current.

In a more general manner, $\tau_m$ can be expressed as $\tau_m = \frac{C}{\bar{g}_L + \bar{g}_h A_h^\infty + \bar{g}_h(V - E_h)\frac{\partial A_h}{\partial V}}$ (see Appendix A, Eq A8). To gain a deeper insight, we made a simple characterization of $\frac{\partial A_h}{\partial V}$ and its dependency with $\tau_h$. For this, we ran simulations in which we injected a ZAP stimulus (see Methods), and we measured simultaneously $\Delta A_h$ and $\Delta V$ from the $A_h - V$ trajectories selected at specific slow ($\omega \approx 5$ Hz) and fast frequencies ($\omega \approx 200$ Hz) (Fig 4). The membrane potential was fixed with an external constant current so that the steady voltage is at V = $-80$ mV and we changed the value of $\tau_h = 100, 200, 500$ and $1000$ ms. Our results showed almost vertical trajectories at low frequencies (Fig 4A), but evolving into horizontal and less round at higher frequencies (as in Fig4 B) (we refer the reader to [29] for a thorough discussion of this phenomenon with this very same model and its implication in subthreshold resonance). This can be interpreted as large $\Delta A_h$ and $\Delta V$ at low frequencies that evolve into smaller $\Delta A_h$ and $\Delta V$ (Fig 4C and 4D). As expected, resonant behavior can be observed in Fig 4C, where for a particular frequency there is an amplification of the voltage response. Finally, Fig. 4E shows the behavior of $\frac{\Delta A_h}{\Delta V}$, increasing when $\tau_h$ decreases by the combined effect of larger $\Delta A_h$ and smaller $\Delta V$ for shorter $\tau_h$.

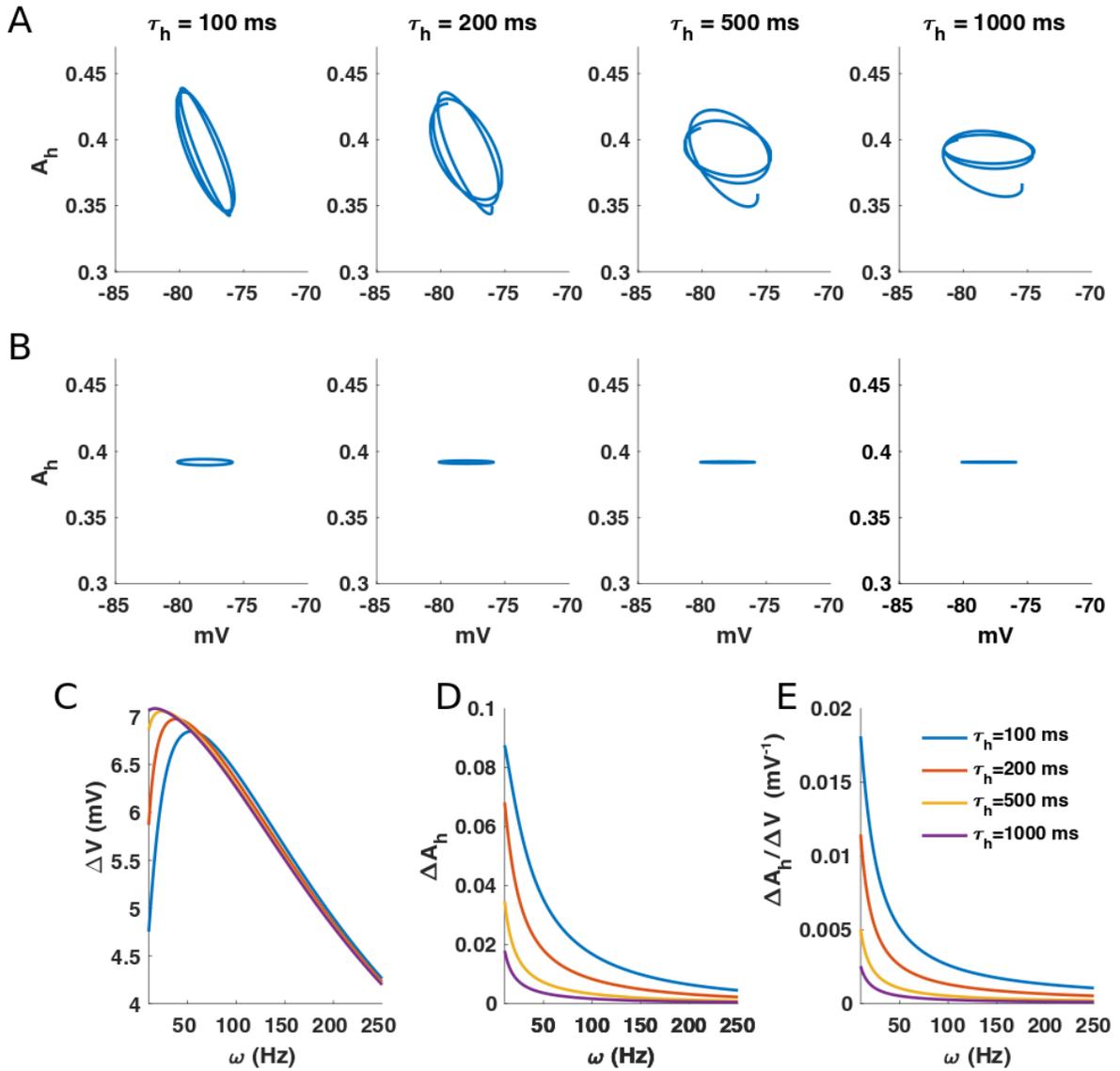

**Fig 4.** Evolution of trajectories for different $\tau_h$. The plots contain activation variable ($A_h$) versus membrane potential of the neuron while being driven by ZAP stimulus for different activation time constants: $\tau_h = 100, 200, 500$ and $1000$ ms. There are different stimulation frequencies ($\omega$): (A) low frequency ($\omega \approx 5$ Hz) and (B)

high frequency ($\omega \approx 200$ Hz). (C) Variation of V ($\Delta V$), (D) variation of $A_h$ ($\Delta A_h$), and (E) $\Delta A_h/\Delta V$ for different values of $\tau_h$.

The partial derivative ($\frac{\partial A_h}{\partial V}$) in the third term of the denominator has well known analytical solution only for the extreme cases when $\tau_h \to \infty$ (infinitely slow kinetics) and $\tau_h = 0$ (instantaneous kinetics), as shown above. In order to obtain an analytical approximation of $\tau_m$ for any $\tau_h$, that fits the behavior observed in Fig 4E, we propose an approximation of $\frac{\partial A_h}{\partial V}$. For this, we assume that a change in the voltage produces a change in the activation variable $A_h$ that evolves in time following an exponential function [32]; $\frac{\partial A_h}{\partial V} = (1 - e^{-\Delta t/\tau_h})\frac{\partial A_h^\infty}{\partial V}$. See from Fig 5 that the change in $I_h$ activation ($\partial A_h$) is limited by the available time to evolve (represented by $\Delta t$). With enough time, $\partial A_h$ can achieve $\partial A_h^\infty$, otherwise $\partial A_h$ is just a portion of $\partial A_h^\infty$. In that sense, the main effect of the interplay between $\Delta t$ and $\tau_h$ on $A_h$ is to partially activate $I_h$ (Fig 5).

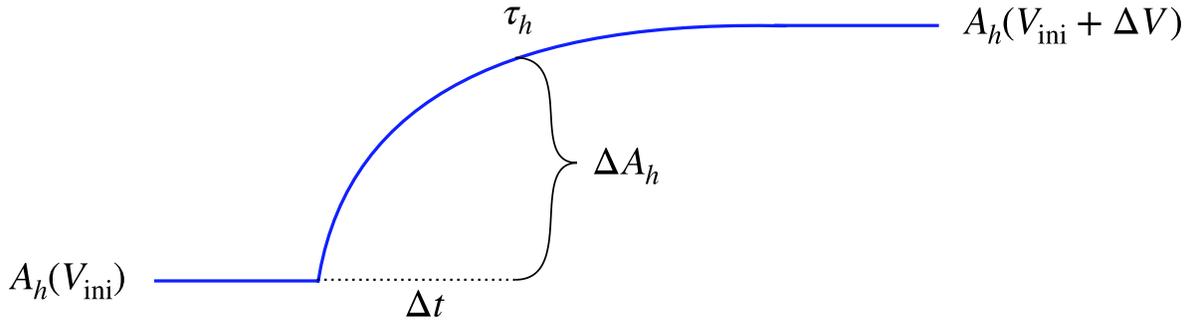

**Fig 5.** Schematic diagram of $\Delta A_h$ evolution over time. For the same $\Delta t$, $\Delta A_h$ is bigger for lower $\tau_h$.

The time interval $\Delta t$ in which changes in $A_h$ have an influence on $\tau_m$ is also $\tau_m$. In a single compartment with a leak current plus $I_h$, the longest $\tau_m$ corresponds to the case when all $I_h$ contributions are neglected, i.e., when $\tau_m = \tau_L = C/\bar{g}_L$. Thus, we approximate the equation above to this case as follows: $\frac{\partial A_h}{\partial V} = (1 - e^{-\tau_L/\tau_h})\frac{\partial A_h^\infty}{\partial V}$. The term

$$\alpha(\tau_L, \tau_h) = \left(1 - e^{-\tau_L/\tau_h}\right), \qquad (8)$$

will be called "time scaling" (or α) factor.. Eq. 8 is extremely valuable since it provides a general understanding about the interaction of the temporal dynamics imposed by the passive current ($\tau_L$) and the $I_h$ kinetics ($\tau_h$) currents as well as how this interaction determines the α factor. To verify its consistency, we check it using two scenarios. The first is when $\tau_h \to 0$, then $e^{-\tau_L/\tau_h} \to 0$, and $\alpha \to 1$. The second is when $\tau_h \to \infty$, then $e^{-\tau_L/\tau_h} \to 1$, and $\alpha \to 0$. Both results were expected, confirming the consistency of the α time scaling factor.

Thus, an approximate solution for $\tau_m$ can be expressed as: $\tau_m = \frac{C}{\bar{g}_L + \bar{g}_h A_h^\infty + \alpha \bar{g}_h (V - E_h)\frac{\partial A_h^\infty}{\partial V}}$, where the term $\alpha \bar{g}_h (V - E_h)\frac{\partial A_h^\infty}{\partial V} = \alpha G_h^{Der}$. Writing the equation in a simpler manner:

$$\tau_m = \frac{C}{\bar{g}_L + g_h + \alpha(\tau_h) G_h^{Der}}. \qquad (9)$$

The α factor is dependent on $\tau_h$ in a way that when $\tau_h \to 0$ we have $\alpha \to 1$ and when $\tau_h \to \infty$ we have $\alpha \to 0$. However, the dependency of the α factor with $\tau_h$, and whether this is the only factor determining it remains unknown. Therefore, in order to gain a deeper insight of the effect of $\tau_h$ on $\tau_m$, we must first characterize the α factor.

This mathematical analysis allows us to conclude that the activation time constant ($\tau_h$) is able to influence $\tau_m$ by modulating the contribution of the derivative conductance $G_h^{Der}$ to the $\tau_m$ by changing the values of the α factor from 0 to 1 when $\tau_h$ changes from infinity to zero.

In the next sections we will use simulations in order to obtain an empirical form for the α factor. This new concept will be useful to reproduce quantitatively the effect of the $I_h$ with different $\tau_h$ constants on $\tau_m$.

### 3.3. The α factor is mostly affected by the leak membrane time constant

As discussed before, $I_h$ has activation time constant values ranging from tens of milliseconds to several seconds. In this regard, $I_h$ cannot be treated as an instantaneous or infinitely slow current. In order to study the effect of $\tau_h$ on $\tau_m$, we must first characterize the α factor. To this end, in this section we will focus on the characterization of the α factor. Thus, we ran simulations of a single compartment with a leak current and $I_h$, and then we vary $\tau_h$ from 20 ms to 1000 ms as we did in Fig 2E.

In order to isolate the values of the α factor, we used the relationship $\tau_m = C/(\bar{g}_L + g_h + \alpha(\tau_h) G_h^{Der})$. The procedure adopted to obtain the values of α is the following: assuming the relationship $\tau_m = C/(\bar{g}_L + g_h + Y(V, \tau_h)G_h^{Der})$, we isolated $Y(V, \tau_h)$ and obtained its values using the values of $\tau_m$ measured from the simulations, the values of $G_h^{Der}$ and $g_h$ were obtained from Eqs. 6 and 7 and $\bar{g}_L = 10$ nS and $C = 154$ pF. Hence, $Y(V, \tau_h) = (C/\tau_m - \bar{g}_L - g_h)/G_h^{Der}$. Although $Y(V, \tau_h)$ clearly displays a voltage dependent relation (Fig 6A), for the sake of simplicity, we approximate the α factor as the averaged value of Y(V), with which we can obtain a voltage independent expression for α (Fig 6B). In that way, the α factor is a nonlinear monotonic decreasing function of $\tau_h$.

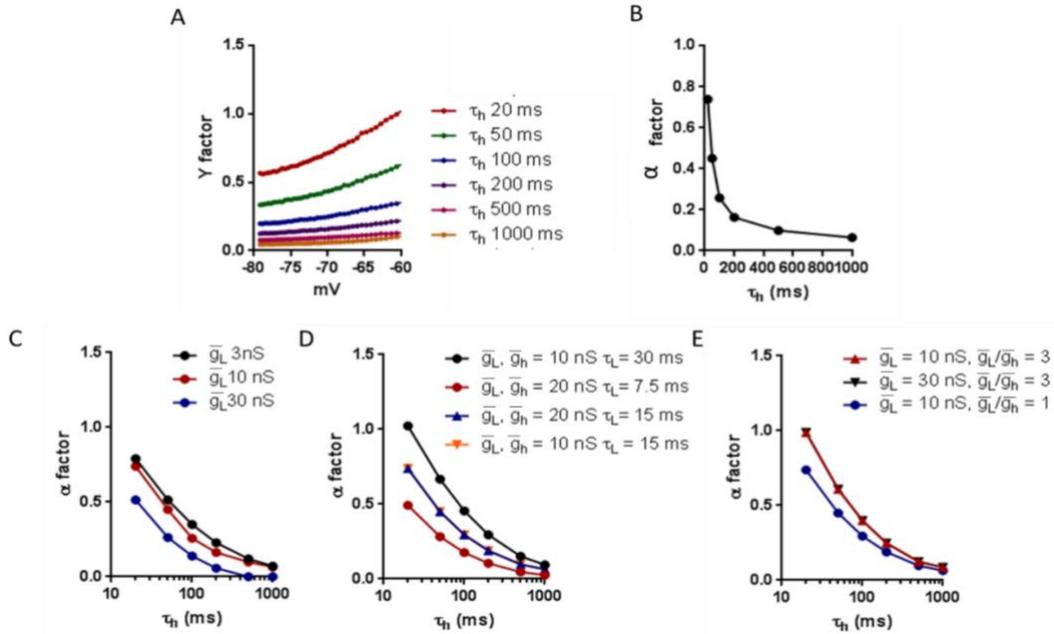

Fig 6. Computational simulations for a single compartment with a leak current and $I_h$ with $\tau_h$ 20, 50, 100, 200, 500 and 1000 ms. A. We calculated the Y(V) values using the relationship $Y(V) = (C/\tau_m - \bar{g}_L - g_h)/G_h^{Der}$. B. α factor values obtained from the relationship α = mean(Y(V)). C. α factor obtained from simulations with $\overline{g_h} = 10$ nS and $\overline{g_L}$ with values 3, 10 and 30 nS. D. α factor obtained from simulations with $\overline{g_L} = \overline{g_h} = 10$ or 20 nS and with $\tau_L$ values 30, 7.5 and 15 mS. E. α factor obtained from simulations with $\overline{g_L} = \overline{g_h} = 3$, 10 or 30 nS and with the ratio $\overline{g_L}/\overline{g_h}$ values 1 and 3.

Intuitively, it is expected that the effects of $\tau_h$ on $I_h$ activation might strongly depend on the temporal filtering imposed by the passive conductance. For instance, for a voltage change, if $\tau_h \gg \tau_m$, the $I_h$ only changes partially but do not reach its steady state before the voltage steady state. In contrast, if $\tau_h \ll \tau_m$, then $I_h$ fully changes, always reaching the steady state [29]. Thus, it seems reasonable to explore the dependency of the α factor with the passive properties. Then, once we isolate the α factor from simulations and characterize its dependency with $\tau_h$, we will be able to further characterize the α factor in conditions when $\overline{g_L}$ changes.

Therefore, we ran simulations of a single compartment with a leak current and $I_h$, with $\tau_h$ values from 20 to 1000 ms and with $\overline{g_L}$ values of 3, 10 and 30 nS.

In the previous section, we showed the analysis conducted in order to obtain the α factor with $\overline{g_L}$ and $\overline{g_h}$ values of 10 nS. Now, in order to determine if the α factor is affected by the value of the leak conductance $\overline{g_L}$, we ran two more simulations using $\overline{g_L}$ = 3 and 30 nS. The curves of the α factor when $\overline{g_L}$ = 3 nS and when $\overline{g_L}$ = 10 nS are similar, but, when $\overline{g_L}$ = 30 nS, the α factor reaches zero for smaller values of $\tau_h$ (Fig 6C). These results suggest that the α factor is a function of $\overline{g_L}$ in a manner that increasing $\overline{g_L}$ accelerates the decay of the α factor. Summarizing, our results suggest that increasing the leak conductance the α factor decreases consequently decreasing the impact of $\tau_h$ on $\tau_m$.

The leak conductance ($\overline{g_L}$) can influence the α factor through different parameters: directly by the $\overline{g_L}$ value, indirectly through the ratio $\overline{g_L}/\overline{g_h}$, or by the membrane time constant of the leak conductance ($\tau_L = C/\overline{g_L}$). We established by which parameter the leak conductance influences the α factor through simulations in which we change $\overline{g_L}$, $\overline{g_h}$, C and $\tau_L$. In these simulations, the values of $\overline{g_L}$ and $\overline{g_h}$ were the same in order to keep constant its ratio, i.e. $\overline{g_L}/\overline{g_h}$ = 1. In total four combinations ($\overline{g_L}$, C, $\tau_L$) were used: (10 nS, 300 pF, 30 ms), (20 nS, 150 pF, 7.5 ms), (20 nS, 300 pF, 15 ms), and (10 nS, 150 pF, 15 ms) (Fig 6D). The results show that whenever $\tau_L$ was fixed to a value (e.g. $\tau_L$ = 15 ms) the α factor obtained was the same, independently of the values of $\overline{g_L}$ or $\overline{g_h}$, (Fig 6D). Moreover, when the $\overline{g_L}$ or $\overline{g_h}$ were fixed, the α factor was different whenever $\tau_L$ was different in a manner that the α factor was lower when $\tau_L$ was lower suggesting a correlation. Our observations suggest that the α factor is mostly dependent on $\tau_L$, but not a function of $\overline{g_L}$ or $\overline{g_h}$.

Finally, we tested the effect of the ratio $\overline{g_L}/\overline{g_h}$ on the α factor. In this case, we simulated changes in the values of $\overline{g_L}$ and $\overline{g_h}$, keeping constant $\tau_L$ = 15 ms (Fig 6E). Next, three combinations ($\overline{g_L}$, $\overline{g_L}/\overline{g_h}$) were used: (10 nS, 1), (10 nS, 3), and (30 nS, 3). The results showed that whenever the ratio $\overline{g_L}/\overline{g_h}$ was fixed (e.g. $\overline{g_L}/\overline{g_h}$ = 3), the α factor obtained was the same, independently of the $\overline{g_L}$ or $\overline{g_h}$ values (Fig 6E). When $\overline{g_L}$ or the $\overline{g_h}$ value was fixed, the α factor was different when the ratio $\overline{g_L}/\overline{g_h}$ was different in a manner that the α factor was lower when the ratio $\overline{g_L}/\overline{g_h}$ was lower, also suggesting a correlation. Our observations suggest that the α factor is a function that depends on the ratio $\overline{g_L}/\overline{g_h}$.

As we have already mentioned, in Fig 6C when $\overline{g_L}$ value increases the α factor decreases. Interestingly, in Fig 6C, the values of $\overline{g_h}$ and C were kept constant, then when the $\overline{g_L}$ value was increased, simultaneously, the ratio $\overline{g_L}/\overline{g_h}$ also increased and $\tau_L$ decreased. However, our previous results (see above) showed that the α factor was lower when $\tau_L$ or the ratio $\overline{g_L}/\overline{g_h}$ was lower. This suggests that $\tau_L$ is a more influential parameter than the ratio $\overline{g_L}/\overline{g_h}$ in the determination of the α factor. Overall, the results in this section are important not only to understand the α factor dependencies but also to determine it from experimental setups.

### 3.4. Numerical predictions of $\tau_m$ using the α factor

In this section we validate our mathematical approximation by comparing it with simulations exploring the values of $\tau_h$ and $\tau_L$. Figure 7 shows values of $\tau_m$ for different combinations of $\tau_h$ (20, 100 and 1000 ms) obtained from simulations and the $\tau_m$ values are predicted using Eqs. 8 and 9. For the combinations ($\overline{g_L}$, $\tau_L$) we used: (3 nS, 45 ms), (10 nS, 15 ms) and (30 nS, 5 ms) (Fig 7A, 7B and 7C). Interestingly, the maximum difference of the $\tau_m$ between the simulated and the theoretical values were 3, 1.17, and 0.3 ms for $\tau_L$ = 45, 15 and 5 ms, respectively, suggesting a good agreement between the approximation and the simulation. We conclude that approximating the α factor as an exponential function with the ratio $\tau_L/\tau_h$ is sufficient to completely reproduce the quantitative tendencies of $\tau_m$ observed in the simulations.

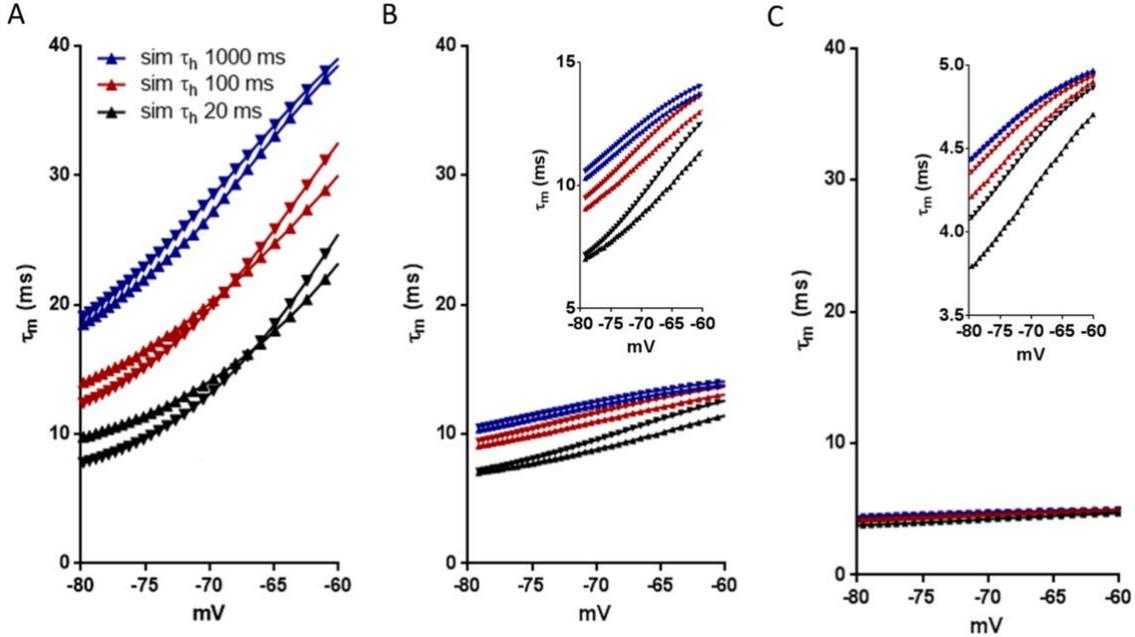

**Fig 7.** Computational simulations of a single compartment with a leak current and $I_h$ with $\tau_h$ 20, 100 and 1000 ms. $\tau_m$ measured from the simulations are the triangle symbols. $\tau_m$ calculated by the relationship $\tau_m = C/(\bar{g}_L + g_h + \alpha(\tau_h) G_h^{Der})$ are the down triangle symbols, where the α factor was determined using $\alpha = 1 - \exp(-\tau_L/\tau_h)$. A. $\tau_m$ from simulations where $\tau_L = 45$ ms. B. The same as A but in this case $\tau_L = 15$ ms. C. The same as A but in this case $\tau_L = 5$ ms.

## 4. DISCUSSION

The aim of this work was to establish the role of hyperpolarization-activated current ($I_h$) kinetics on the membrane time constant ($\tau_m$) of a neuron model, and on the shape of excitatory postsynaptic potentials (EPSPs) in this neuron. Our simulations showed that $I_h$ with faster activation rates attenuates and shortens the EPSPs more efficiently than $I_h$ with slower activation rates. Since one of the most important physiological effects of $\tau_m$ in neurons is the shape of the EPSPs, especially on its descending phase, we demonstrated that the influence of $\tau_h$ on $\tau_m$ is responsible for the influence of $\tau_h$ on the EPSP decay.

Accordingly, it has been demonstrated that an increase in conductance leads to a decrease in $\tau_m$ following a shortening of EPSPs which mostly affects its decay phase [33]. In addition, experimental and computational studies have shown that $I_h$ mainly affects the decay phase of EPSPs rather than the rise phase [34, 35]). And it is well known that in simple point-neuron models, the EPSP decay phase is mostly determined by $\tau_m$ [36]. Moreover, we further confirmed that changes in $\tau_m$ are correlated with changes in EPSPs temporal properties, regardless of the source of these variations (e.g. changes in conductance, capacitance and $I_h$ kinetics). Overall, we provide compelling evidence that $I_h$ kinetics modulates the EPSP shape by its influence upon $\tau_m$ confirming a longstanding empirical observation.

It is well known that $I_h$ can control membrane input resistance and the resting membrane potential (RMP) as well as the membrane time constant [37,38]. Our results suggest that $I_h$ kinetics selectively modulates $\tau_m$ without affecting other neuronal properties such as the resting membrane potential and membrane input resistance, which are only determined by the $I_h$ conductance. Thus, variations in the $I_h$ activation/deactivation kinetics can selectively modulate the EPSPs shape and temporal properties of the neuron with undetected impact on other subthreshold properties (e.g. membrane input resistance or resting membrane potential).

In order to elucidate the biophysical mechanism underlying the influence of $\tau_h$ on $\tau_m$, we proposed a new kinetics-dependent analytical solution of $\tau_m = C/(\bar{g}_L + g_h + \alpha(\tau_h) G_h^{Der})$, where the α factor is dependent on $\tau_h$ and $\tau_L$. This last relationship establishes that the $I_h$ derivative conductance $G_h^{Der}$ contributes controlling the $\tau_m$, although this control is limited by the α factor. In a previous work, Drion and colleagues studied the dynamical interplay among many different ionic currents in distinct timescales using the analysis of dynamic input conductances [39]. The mathematical derivations in their work are conceptually similar to ours, mostly regarding the derivation of the α factor (called "voltage-dependent weighting factor" in [39]). Although we obtained a different mathematical expression, [39], the voltage-dependent weighting factor was defined as logarithmic distances among activation time constants whereas in ours we defined the α factor as an exponential function of the ratio $\tau_L/\tau_h$, we believe these definitions are mechanistically similar. Drion and colleagues used this approach to investigate the mechanisms of ion channel interplay in shaping suprathreshold activity (spikes), whereas we investigated the shaping of subthreshold activity (EPSPs and $\tau_m$).

Importantly, we get a good agreement between the theoretical and simulated data using an approximation of the α factor that follows as an exponential function of the ratio $\tau_L/\tau_h$. The α factor is a nonlinear monotonic decreasing function of $\tau_h$ and an increasing function of $\tau_L$. Then, decreasing $\tau_h$ increases the α factor values and consequently increases the impact of $\tau_h$ on $\tau_m$. Moreover, increasing the leak conductance decreases $\tau_L$ which in turn decreases the α factor values and consequently decreases the impact of $\tau_h$ on $\tau_m$.

Our theoretical predictions can be tested in real neurons using the dynamic clamp technique as being reported elsewhere [28,37]. For this, it is possible to study the effect of the $I_h$ kinetics on neuron excitability injecting an artificial $I_h$ in neurons using the Eq. 2, Eq. 3, and Eq. 4, and the values for maximum conductance $\bar{g}_h$, the reversal potential $E_h$ and the activation time constant $\tau_h$ (fast and slow, respectively) which can be obtained from previous reports [37]. Furthermore, the α factor can be easily determined from voltage-clamp experiments by measuring the leak and $I_h$ time constants ($\tau_L$ and $\tau_h$)..

Our results should also be valid and extended for multiple $I_h$ currents. In fact, each current will have its own α factor determined from the set of $\tau_L$ and $\tau_h$ (see [40] for an example of experiments). Such an extension may give rise to new interpretations of how a neuron performs temporal integration of an input under the influence of multiple ionic current.

In conclusion, our results expand the membrane time constant definition from the classical passive cable equations and should be useful for further interpretations of processing employed by neurons when receiving synaptic inputs.

## Acknowledgements

This article was produced as part of the IRTG 1740/TRP 2011/50151-0, funded by the DFG/FAPESP. It was also supported partially by the S. Paulo Research Foundation (FAPESP) Research, Innovation and Dissemination Center for Neuromathematics (CEPID NeuroMat, Grant No. 2013/07699-0). The authors also thank FAPESP support through Grants Nos. 2013/25667-8 (R.F.O.P.), 2015/50122-0 and 2018/20277 (A.C.R.). C.C.C. was supported by a CAPES PhD scholarship. A.C.R. thanks financial support from the National Council of Scientific and Technological Development (CNPq), Grant No. 306251/2014-0. This study was financed in part by the Coordenação de Aperfeiçoamento de Pessoal de Nível Superior-Brasil (CAPES) - Finance Code 001.

## Author contribution statement

Author Contributions: C.C.C conceived the work, C.C.C. and R.F.O.P. run and analyze the simulations, C.C.C. and R.F.O.P and A.C.R. wrote the manuscript, A.C.R. supervised the research. All authors read and agreed to the published version of the manuscript.

**Appendix A. Membrane time constant is modulated by rate of activation of $I_h$**

A single compartment neuron with one leak current ($I_L$) and one $I_h$ has the membrane equation:

$$C\frac{dV}{dt} = -I_L - I_h \tag{A1}$$

And assuming that the voltage temporal evolution of the capacitor charging can be described by a single exponential function when a constant current step is injected [32]:

$$V(t) = B\left(1 - \exp\left(-\frac{t}{\tau_m}\right)\right) + V_0, \tag{A2}$$

where B and $V_0$ are constants and $\tau_m$ is the membrane time constant. Then differentiating equation (A2) in time and multiplying by C:

$$C\frac{dV}{dt} = CB\left(\frac{e^{-\frac{t}{\tau_m}}}{\tau_m}\right) \tag{A3}$$

Isolating the exponential term in Eq A2 we get that $e^{-\frac{t}{\tau_m}} = 1 - \frac{V(t)-V_0}{B}$, and substituting this in Eq (A3)

$$C\frac{dV}{dt} = CB\left(\frac{1 - \frac{V(t)-V_0}{B}}{\tau_m}\right) \tag{A4}$$

Equaling A4 and A1 and differentiating with respect to V:

$$\frac{C}{\tau_m} = \frac{\partial I_L}{\partial V} + \frac{\partial I_h}{\partial V} \tag{A5}$$

Isolating $\tau_m$:

$$\tau_m = \frac{C}{\frac{\partial I_L}{\partial V} + \frac{\partial I_h}{\partial V}} \tag{A6}$$

Differentiating each current, we obtain:

$$\tau_m = \frac{C}{\bar{g}_L + \overline{g_h}A_h + \overline{g_h}(V-E_h)\frac{\partial A_h}{\partial V}} \tag{A7}$$

For the sake of simplicity, we will only investigate the cases where $\Delta A_h$ is small due to small $\Delta V$. Under this condition, $\overline{g_h}A_h \approx \overline{g_h}A_h^\infty$, where this term corresponds to the $I_h$ chord conductance ($g_h$). Replacing in the equation:

$$\tau_m = \frac{C}{\bar{g}_L + \overline{g_h}A_h^\infty + \overline{g_h}(V-E_h)\frac{\partial A_h}{\partial V}} \tag{A8}$$

The partial derivative ($\frac{\partial A_h}{\partial V}$) in the third term of the denominator has well known analytical solution only for the extreme cases when $\tau_h \to \infty$ (infinitely slow kinetics) and $\tau_h = 0$ (instantaneous kinetics). When $\tau_h \to \infty$, $A_h$ does not change with V, then $A_h = A_h^\infty(V_0)$ and $\frac{\partial A_h}{\partial V} = 0$, then:

$$\tau_m(\tau_h \to \infty) = \frac{C}{\bar{g}_L + \overline{g_h}A_h^\infty} = \frac{C}{\bar{g}_L + g_h} \tag{A9}$$

On the other hand, when $\tau_h = 0$, $\frac{\partial A_h}{\partial V} = \frac{\partial A_h^\infty}{\partial V}$, then:

$$\tau_m = \frac{C}{\bar{g}_L + \bar{g}_h A_h^\infty + \bar{g}_h(V - E_h)\frac{\partial A_h^\infty}{\partial V}} = \frac{C}{\bar{g}_L + G_h} \tag{A10}$$

Where $G_h$ is the slope conductance ($G_h$). Concluding, we can state that $\tau_m$ is determined by the steady state slope conductance of the instantaneous current and chord conductance of the infinitely slow current.